\documentstyle[12pt]{article}
\def\beq{\begin{equation}}
\def\eeq{\end{equation}}
\def\bea{\begin{array}}
\def\eea{\end{array}}
\def\beqa{\begin{eqnarray}}
\def\eeqa{\end{eqnarray}}
\newcommand{\refeq}[1]{\mbox{eq.~(\ref{eq:#1})}}
\newcommand{\half}{{\scriptstyle{{1\over 2}}}}
\newcommand{\quarter}{{\scriptstyle{{1\over 4}}}}
\def\myIm{{\Im m}}

\def\ie{{i.e.\/}}\relax
\relax
\relax
\newcommand{\zahlen}{{\rm Z \!\! Z}}
\newcommand{\re}{\relax{\rm I\kern-.18em R}}

\def\cT{{\cal{T}}}

\def\cP{{\cal{P}}}
\def\ddz{{\frac{d}{dz}}}
\def\Tr{{\rm Tr}}
\def\tr{{\rm tr}}
\def\pl{{{\cal P}_\infty}}
\def\plo{{{\cal P}_\infty^0}}
\def\SD{self\--du\-al}
\def\ASD{an\-ti\--\-self\-du\-al}

\begin{document}
\hfill INLO-PUB-9/98
\vskip1cm
\begin{center}
{\LARGE{\bf{\underline{Monopole Constituents inside}}}}\\
{\LARGE{\bf{\underline{SU(n) Calorons}}}}\\
\vspace{7mm}
{\large Thomas C. Kraan and Pierre van Baal} \\
\vspace{5mm}
Instituut-Lorentz for Theoretical Physics, University of Leiden,\\
PO Box 9506, NL-2300 RA Leiden, The Netherlands.
\end{center}
\vspace*{5mm}{\narrower\narrower{\noindent
\underline{Abstract:}
We present a simple result for the action density of the $SU(n)$ charge one 
periodic instantons - or calorons - with arbitrary non-trivial Polyakov loop 
$\pl$ at spatial infinity. It is shown explicitly that there are $n$ lumps 
inside the caloron, each of which represents a BPS monopole, their masses 
being related to the eigenvalues of $\pl$. A suitable combination of the 
ADHM construction and the Nahm transformation is used to obtain this result.
}\par}
\section{Introduction}
Instantons and BPS monopoles are self-dual finite action solutions to the 
Yang-Mills equations of motion. Their origin is topological and related to 
windings in the gauge transformations describing their behaviour at infinity, 
where these solutions approach vacua, necessary for the action to be finite. 
The action of these solutions is proportional to these winding numbers or 
charges, which suggests that they are composed out of elements with unit 
charge. We will find by computing the action density that even charge one 
periodic instantons (calorons) for the gauge group $SU(n)$ are further 
composed out of constituents. The constituents are $n$ basic BPS 
monopoles~\cite{BPS}, whose magnetic charges cancel exactly. 

Periodic instantons were first discussed in the context of finite temperature 
field theory~\cite{HarShe,GroPisYaf}. Motivated by issues of 
T-duality~\cite{PLB} and D-brane constructions~\cite{LeeYi1} in 
string theory, periodic instantons with a non-trivial Polyakov loop were 
recently constructed. The ingredients of the Nahm transformation for $SU(n)$ 
calorons~\cite{NahMonCal}, can be naturally presented in terms of $n$ monopole 
constituents~\cite{GarMur}. This has been the basis of the approach followed 
by Lee and co-workers~\cite{LeeLu,LeeYi2}. Our work relies on a more direct 
approach, combining the ADHM construction of multi-instanton 
solutions~\cite{AtiDriHitMan} with the Nahm construction~\cite{NahMonCal}, 
related by Fourier transformation. The constituent monopoles appear as 
explicit lumps in the action density~\cite{NPB}.

The composite nature of the periodic instantons can also be appreciated by an 
argument due to Taubes~\cite{Taubes} on how to build out of monopoles, 
configurations with non-trivial topological charge. It has far reaching 
consequences, which go beyond the existence of exact caloron 
solutions~\cite{NPB}. Its relevance for QCD has motivated us 
to extend our work to $SU(3)$.

Finite action solutions on $\re^3 \times S^1$ should approach vacua at spatial
infinity. Due to the topology of this base manifold, these vacua can be 
non-trivial.  This endows the caloron with extra parameters - labeling these 
vacua - which are studied in terms of the eigenvalues of the Polyakov loop 
around $S^1$ at spatial infinity. The Polyakov loop is defined in the periodic 
gauge, $A_\mu(\vec x,x_0+\cT)=A_\mu(\vec x,x_0)$, as
\beq
{\cal P} ( \vec x) = P \exp( \int_0^{\cal T} A_0(\vec x, x_0) d x_0),
\eeq
where $\cT$ is the period and $P$ denotes path-ordering. At infinity the 
value of the Polyakov loop does not change under continuous deformations 
of the loop and its eigenvalues are topological invariants.

In extending our work to the $SU(3)$ case it turned out to be natural to 
generalise to $SU(n)$ (see also the appendices of ref.~\cite{LeeYi2}). We 
will present the formula for the action density in section 2. The derivation 
is outlined in section 3. A detailed description will be given elsewhere. 
In section 4 we discuss the properties of the solution.

\section{The result}
We consider the calorons with no net magnetic charge, in which case the 
Polyakov loop (holonomy) at spatial infinity becomes constant. Its eigenvalues 
$e^{2\pi i\mu_m}$ will play an important role in the construction,
\beq
\lim_{|\vec x|\rightarrow\infty}{\cal P}(\vec x)=\pl=V\plo V^{-1},\quad\plo
=\exp[2\pi i{\rm diag}(\mu_1,\ldots,\mu_n)].
\eeq
Making use of the gauge symmetry, we can choose the eigenvalues such that
\beq
\mu_1<\ldots<\mu_n<\mu_{n\!+\!1}\equiv\mu_1+1,\quad\sum_{m=1}^n\mu_m=0,
\eeq
assuming maximal symmetry breaking for the moment. We define $\nu_m=\mu_{m\!+
\!1}-\mu_m$, related to the mass of the $m^{\rm th}$ constituent 
monopole. Standard arguments, summarised below, gives $4n$ instanton 
parameters for fixed $\pl$, including the global gauge transformations 
that do not change $\pl$. We will see that $3n$ parameters can be interpreted 
as the positions ($\vec y_m$) of the constituents. The remaining parameters 
in this interpretation are the $n-1$ phases related to the unbroken gauge group
$U(1)^{n-1}$, on which the action density does not depend and the position of 
the caloron in time, which we fix to be 0 by translational invariance. Also 
we will use the scale invariance to set $\cT=1$. Where needed, the proper 
$\cT$ dependence can be reinstated on dimensional grounds. We find the 
following surprisingly simple formula
\beq
- \half {\rm tr} F_{\mu\nu}^2 = - \half \partial_\mu^2 \partial_\nu^2 \log
\psi.
\label{eq:adpsi}
\eeq
where the positive scalar potential $\psi$ is defined as
\beq
\psi(x)=\half\tr\prod_{m=1}^n\left\{\left(\!\!\!\bea{cc}r_m&|\vec y_m\!-
\!\vec y_{m+1}|\\0&r_{m+1}\eea\!\!\!\right)\left(\!\!\!\bea{cc}\cosh(2\pi\nu_m
r_m)&\sinh(2\pi\nu_mr_m)\\\sinh(2\pi\nu_mr_m)&\cosh(2\pi\nu_mr_m)\eea\!\!\!
\right)\frac{1}{r_m}\!\right\}\!-\!\cos(2\pi x_0).\label{eq:psi}
\eeq
Here $r_m=|\vec x-\vec y_m|$ denotes the center of mass radius of the
$m^{\rm th}$ monopole. The order of matrix multiplication is crucial,
$\prod_{m=1}^n A_m \equiv A_n \ldots A_1$.

\section{The construction}
In our description of the caloron with non-trivial Polyakov loop, we pick the
so-called algebraic gauge,
\beq
A_\mu(\vec x,x_0+\cT)=\pl A_\mu(\vec x,x_0)\cP^{-1}_\infty,\label{eq:ag}
\eeq
which is related to the periodic gauge by the non-periodic gauge transformation
$g(\vec x, x_0) =V\exp[2 \pi i x_0{\rm diag}(\mu_1,\ldots, \mu_n)/\cT]V^{-1}$. 
In the algebraic gauge {\em all} gauge field components approach zero at 
infinity. The technique we use is to interpret the ADHM data as the Fourier 
coefficients of the functions that appear in the Nahm transformation. This is 
to solve the quadratic ADHM constraint, which is non-trivial for a periodic 
array of instantons, twisted in colour space going from one time slice to the 
next.

We summarise the ADHM formalism for $SU(n)$ charge $k$ 
instantons~\cite{AtiDriHitMan}, to fix our notation. It employs a $k$ 
dimensional vector $\lambda=(\lambda_1,\ldots,\lambda_k)$, where 
$\lambda_i^\dagger$ is a two-component spinor in the $\bar n$ representation of
$SU(n)$. Alternatively, $\lambda$ can be seen as an $n\times 2k$ complex matrix.
In addition one has four complex hermitian $k\times k$ matrices $B_\mu$, 
combined into a $2k\times2k$ complex matrix $B=\sigma_\mu\otimes B_\mu$, using 
the unit quaternions $\sigma_\mu=(1_2,i\vec\tau)$ and $\bar\sigma_\mu=(1_2,-i
\vec\tau)$, where $\tau_i$ are the Pauli matrices. With abuse of notation, we 
often write $B=\sigma_\mu B_\mu$. Together $\lambda$ and $B$ constitute the 
$(n + 2 k)\times 2k$ dimensional matrix $\Delta(x)$, to which is associated 
a complex $(n+2k)\times n$ dimensional normalised zero mode vector $v(x)$,
\beq
\Delta(x)=\left(\!\!\bea{c}\lambda\\B(x)\eea\!\!\right),\quad B(x)=B-x,\quad
\Delta^\dagger(x)v(x)=0,\quad v^\dagger(x)v(x)=1_n.
\eeq
Here the quaternion $x=x_\mu\sigma_\mu$ denotes the position (a $k\times k$ 
unit matrix is implicit) and $v(x)$ can be solved explicitly in terms of the
ADHM data by
\beq
v(x)=\left(\!\bea{c}-1_n\\u(x)\eea\!\right)\phi^{-\half},\quad u(x)=(B^\dagger
-x^\dagger)^{-1}\lambda^\dagger,\quad\phi(x)=1_n+u^\dagger(x)u(x),
\eeq
As $\phi(x)$ is an $n\times n$ positive hermitian matrix, its square root 
$\phi^{\half}(x)$ is well-defined. The gauge field is given by
\beq
A_\mu(x)=v^\dagger(x)\partial_\mu v(x)=\phi^{-\half}(x)(u^\dagger(x)
\partial_\mu u(x))\phi^{-\half}(x)+\phi^{\half}(x)\partial_\mu\phi^{-\half}(x).
\label{eq:aadhm}
\eeq

For $A_\mu(x)$ to be a self-dual connection, $\Delta(x)$ has to satisfy the 
quadratic ADHM constraint, which states that $\Delta^\dagger(x)\Delta(x)=
B^\dagger(x)B(x)+\lambda^\dagger\lambda$ (considered as $k\times k$ complex 
quaternionic matrix) has to commute with the quaternions, or equivalently
\beq
\Delta^\dagger(x)\Delta(x)=\sigma_0\otimes f^{-1}_x,\label{eq:adhmconstr}
\eeq
defining $f_x$ as a hermitian $k\times k$ Green's function. The \SD ity 
follows by computing the curvature
\beq
F_{\mu\nu}=2\phi^{-\half}(x)u^\dagger(x)\eta_{\mu\nu}f_x u(x)\phi^{-\half}(x),
\eeq
making essential use of the fact that $f_x$ commutes with the quaternions, and
$\eta_{\mu\nu}\equiv\sigma_{[\mu}\bar\sigma_{\nu]}$ being \SD\ ($\bar\eta_{\mu
\nu}\equiv\bar\sigma_{[\mu}\sigma_{\nu]}$ is \ASD). The quadratic constraint
can be formulated as $\myIm(\Delta^\dagger(x)\Delta(x))=0$, where
$\myIm W\equiv\half[W-\tau_2 W^t\tau_2]$, and one obtains
\beq
\bar\eta_{\mu\nu}\otimes B_\mu B_\nu+\half\tau_a
\otimes\tr_2(\tau_a\lambda^\dagger\lambda)= 0,\label{eq:radhmcon}
\eeq
where $\tr_2$ is the spinorial trace. Note that this implies that 
$\tr_2(\tau_a\lambda^\dagger\lambda)$ vanishes on the diagonal for $a=1,2,3$. 
To count the number of instanton parameters we observe that the transformation 
$\lambda\rightarrow\lambda T^\dagger$, $B_\mu\rightarrow T B_\mu T^\dagger$,
with $T \in U(k)$ leaves the gauge field and the ADHM constraint untouched. 
Taking this symmetry into account, we find the dimension of the instanton 
moduli space to be $4kn$ dimensional. Global gauge transformations are 
realised by $\lambda \rightarrow g \lambda$, with $g\in SU(n)$. Those that
leave $\pl$ invariant reduce the dimension of the gauge invariant parameter
space (by $n-1$ for maximal symmetry breaking). Finally, we quote an elegant 
result~\cite{Osb} for the action density in terms of $f_x$
\beq
\tr F^2_{\mu\nu}(x)=-\partial^2_\mu\partial^2_\nu\log\det f_x\label{eq:acdens}.
\eeq

The charge one caloron with Polyakov loop $\pl$ at infinity is built out of a
periodic array of instantons, twisted by $\pl$. This is implemented in
the ADHM formalism by requiring (suppressing colour and spinor indices)
\beq
u_{p+1}(x+1)=u_p(x)\cP_\infty^{-1}
\label{eq:ucocyc}
\eeq
with $p\in\zahlen$. Using that $\phi^{\pm\half}(x+1)=\pl\phi^{\pm\half}(x)
\cP_\infty^{-1}$, \refeq{aadhm} leads to the required result, \refeq{ag}. 
Demanding
\beq
\lambda_{p+1}=\pl\lambda_p,\quad B_{p,q}=B_{p-1,q-1}+\sigma_0\delta_{p,q},
\eeq
suitably implements \refeq{ucocyc} and is partially solved by imposing 
\beq
\lambda_p=\cP^p_\infty\zeta, \quad B_{p,q}=p\sigma_0\delta_{p,q}+\hat A_{p-q},
\eeq
with $\hat A$ still to be determined to account for \refeq{radhmcon}, which
also constrains the spinor $\zeta^\dagger$ in the $\bar n$ representation of 
$SU(n)$ to
\beq
\tr_2(\tau_a\zeta^\dagger\zeta)=0,\quad a=1,2,3.
\label{eq:zetacon}
\eeq
It is useful to introduce the $n$ projectors $P_m=VP_m^{(0)}V^{-1}$, 
with $(P_m^{(0)})_{a,b}=\delta_{m,a}\delta_{m,b}$ and
$P_mP_{m'}=\delta_{m,m'}P_m$, such that $\pl=\sum_m\exp(2\pi i\mu_m)P_m$ and
$\lambda_p=\sum_m\exp(2\pi ip\mu_m)P_m\zeta$.

We now perform the Fourier transformation to the Nahm setting~\cite{NahMonCal},
which casts $B$ into a Weyl operator and $\lambda^\dagger\lambda$ into a 
singularity structure on $S^1$,
\beqa
&&\sum_{p,q} B_{p,q}(x) e^{2\pi i(pz-qz')}=\frac{\delta(z-z')}{2\pi i}\hat 
D_x(z'),\quad\hat D_x(z)=\sigma_\mu\hat D_x^\mu(z)=\ddz+\hat A(z)-2\pi ix, 
\nonumber\\ 
&&\hat A(z)=\sigma_\mu\hat A^\mu(z),\quad\hat A^\mu(z)=2\pi i\sum_p
e^{2\pi ipz}\hat A^\mu_{p},\\
&&\sum_p e^{-2\pi piz}\lambda_p=\sum_p e^{2\pi ip(\mu_m-z)}P_m\zeta=\hat\lambda
(z),\quad\hat\lambda(z)=\sum_m\delta(z-\mu_m)P_m\zeta,\nonumber\\
&&\sum_{p,q}\lambda^\dagger_p e^{2\pi i(pz-qz')}\lambda^{\phantom{\dagger}}_q=
\delta(z-z')\hat\Lambda(z),\quad\hat\Lambda(z)=\sum_m\delta(z-\mu_m)
\zeta^\dagger P_m\zeta=\zeta^\dagger\hat\lambda(z).\nonumber
\eeqa
Introducing the vector $\zeta_{(m)}=P_m\zeta$, it is standard to show 
\beq
\zeta^\dagger P_m\zeta=\zeta^\dagger_{(m)}\zeta^{\phantom{\dagger}}_{(m)}=
\frac{1}{2\pi}(|\vec\rho_m|-\vec\tau\cdot\vec\rho_m),
\eeq
and the quadratic ADHM constraint, which takes the form
\beq
\half[\hat D_\mu(z),\hat D_\nu(z)]\bar\eta_{\mu\nu}=4\pi^2\myIm\hat\Lambda(z),
\eeq
leads to the (for $k=1$ abelian) Nahm equation 
\beq
\ddz\hat A_j(z)=2\pi i\sum_m\delta(z-\mu_m)\rho_m^j. \label{eq:abelnahm}
\eeq
The $T$ symmetry in the ADHM construction translates into a $U(1)$ gauge 
symmetry on $S^1$, which leaves $\hat A_i$ invariant and allows one to set 
$\hat A_0=2\pi i\xi_0$, $\xi_0$ being the position in time, which we absorb in 
$x_0$. Since $\vec\rho_m=-\pi\tr_2(\vec\tau\zeta^\dagger P_m\zeta)$, it follows 
that $\sum_m\vec\rho_m=\!-\pi\tr_2(\vec\tau\zeta^\dagger\zeta)\!=\!\vec 0$, see 
\refeq{zetacon}, such that we may introduce $\vec y_m$ ($\vec y_0\equiv
\vec y_n$), with $\vec\rho_m=\vec y_m-\vec y_{m-1}$. In terms of these we find
\beq
\hat A_j(z)=2\pi i\sum_m\chi_{[\mu_m,\mu_{m+1}]}(z)y_m^j,
\eeq
where $\chi_{[\mu_m,\mu_{m+1}]}(z)=1$ for $z\in[\mu_m,\mu_{m+1}]$ and 0 
elsewhere, taking into account $z$ has period 1. Note that $\vec y_m$ is 
only fixed up to a constant $\vec\xi$, related to the freedom of adding
a constant to the solution of the Nahm equation, \refeq{abelnahm}. The
4-vector $\xi_\mu$ describes the position of the caloron. Also note that 
the $\vec\rho_m$ are independent of the phases of $P_m\zeta$, affected 
by the residual gauge symmetry, of which $n-1$ are independent due to 
the gauge group being $SU(n)$ rather than $U(n)$.

The vector $\vec y_m$ is interpreted as the position of the  $m^{\rm th}$ 
constituent. On each sub-interval $[\mu_m,\mu_{m+1}]$, $\hat A_j(z)=
2\pi iy^j_m$ is constant, which is precisely the Nahm datum for a single BPS 
monopole located at $\vec y_m$~\cite{Nahm}. The length of the Nahm interval 
for the single BPS monopole corresponds to its asymptotic Higgs value, and 
thereby to its mass. Thus, the $m^{\rm th}$ subinterval $[\mu_m,\mu_{m+1}]$ 
corresponds to a BPS monopole at $\vec y_m$, with mass proportional to
$\nu_m=\mu_{m+1}\!-\!\mu_m$. Together, the $n$ BPS monopoles form the $SU(n)$ 
caloron.

The Green's function $f_x$, central in the ADHM construction, is found after a
Fourier transformation of \refeq{adhmconstr}, introducing $\hat f_x(z,z')\equiv
\sum_{p,q}f_x^{p,q}e^{2\pi i(pz-qz')}$ as the solution to the differential 
equation 
\beq
\left\{\!\!\left(\!\frac{1}{2\pi i}\ddz\!-\!x_0\right)^{\!\!2}+\sum_m\chi_{[
\mu_m,\mu_{m+1}]}(z)\,r_m^2+\frac{1}{2 \pi}\sum_m \delta(z\!-\!\mu_m)|\vec y_m
\!-\!\vec y_{m-1}|\!\right\}\!\!\hat f_x(z,z')=\delta(z-z'),\label{eq:df}
\eeq
where the radii are given by $r_m=|\vec x\!-\!\vec y_m|$, to be interpreted as 
the center of mass radii of the constituent monopoles. The solution of a 
quantum-mechanical problem on the circle with a piecewise constant potential 
and delta function impurities is obtained by solving it on each sub-interval, 
where $\hat f_x(z,z')$ is of simple exponential form. Starting at $z=z'$ and 
matching properly at $z=\mu_m$ so as to account for the scattering by the 
impurity, we can go full circle to return at $z=z'$ where one last matching 
accounts for the delta function at the rhs. of \refeq{df}.

With the solution for $f_x(z,z')$ available, we can compute $\Tr\hat D_x^\mu
f_x$ and show it to be equal to $-\pi i\partial_\mu\log\psi$, $\psi$ being 
the positive scalar function defined in \refeq{psi}. Using \refeq{acdens}, 
this leads to \refeq{adpsi}. One retrieves our $SU(2)$ results of 
refs.~\cite{PLB,NPB} by putting $\mu_1=-\omega$, $\mu_2=\omega$ and $\mu_3
=1-\omega$, such that $\nu_1=\mu_2-\mu_1=2\omega$ and $\nu_2=\mu_3-\mu_2=
1-2\omega\equiv2\bar\omega$. Furthermore we identify $r_1=s$, $r_2=r$ and 
$|\vec y_1-\vec y_0|=|\vec y_2-\vec y_1|=\pi\rho^2$.

\section{Discussion}
We briefly discuss the properties of the $SU(n)$ charge one caloron. 
From \refeq{psi} we see that the $m^{\rm th}$ constituent monopole can 
be located at arbitrary $\vec y_m$, with arbitrary mass $8\pi^2\nu_m/\cT$, 
subject only to the constraint $\sum_m\nu_m=1$, choosing $\pl$, 
$\vec\xi$ and $\zeta$ appropriately. As the action density, \refeq{adpsi}, 
is expressed as a total derivative, the action is easily found by 
partial integration, with the expected result of $8\pi^2$.
The size of the instanton is related to the differences 
in position of the constituent monopoles. As we work in units of $\cT$, 
the situation of a small scale (nearby constituents), corresponds to 
large $\cT$, \ie\ to an instanton on $\re^4$. At the other extreme one has
well separated lumps for small $\cT$, \ie\ in the static limit. 
In figure 1 we present a typical $SU(3)$ caloron for decreasing values 
of $\cT$ using \refeq{adpsi}. We will resist the temptation of showing 
results for other $n$, as \refeq{adpsi} can be readily implemented.

\begin{figure}[htb]
\vspace{9.8cm}
\includegraphics{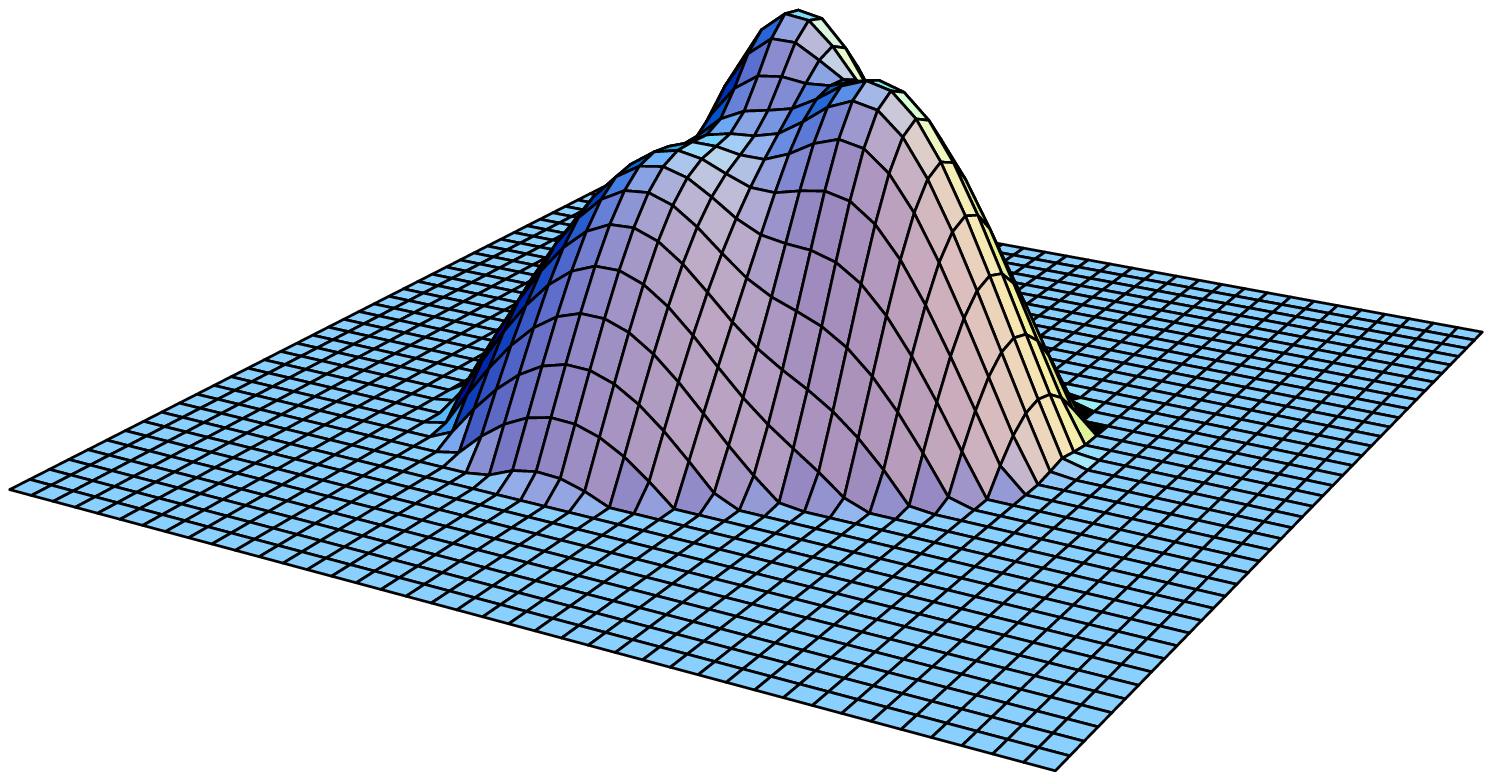}
\includegraphics{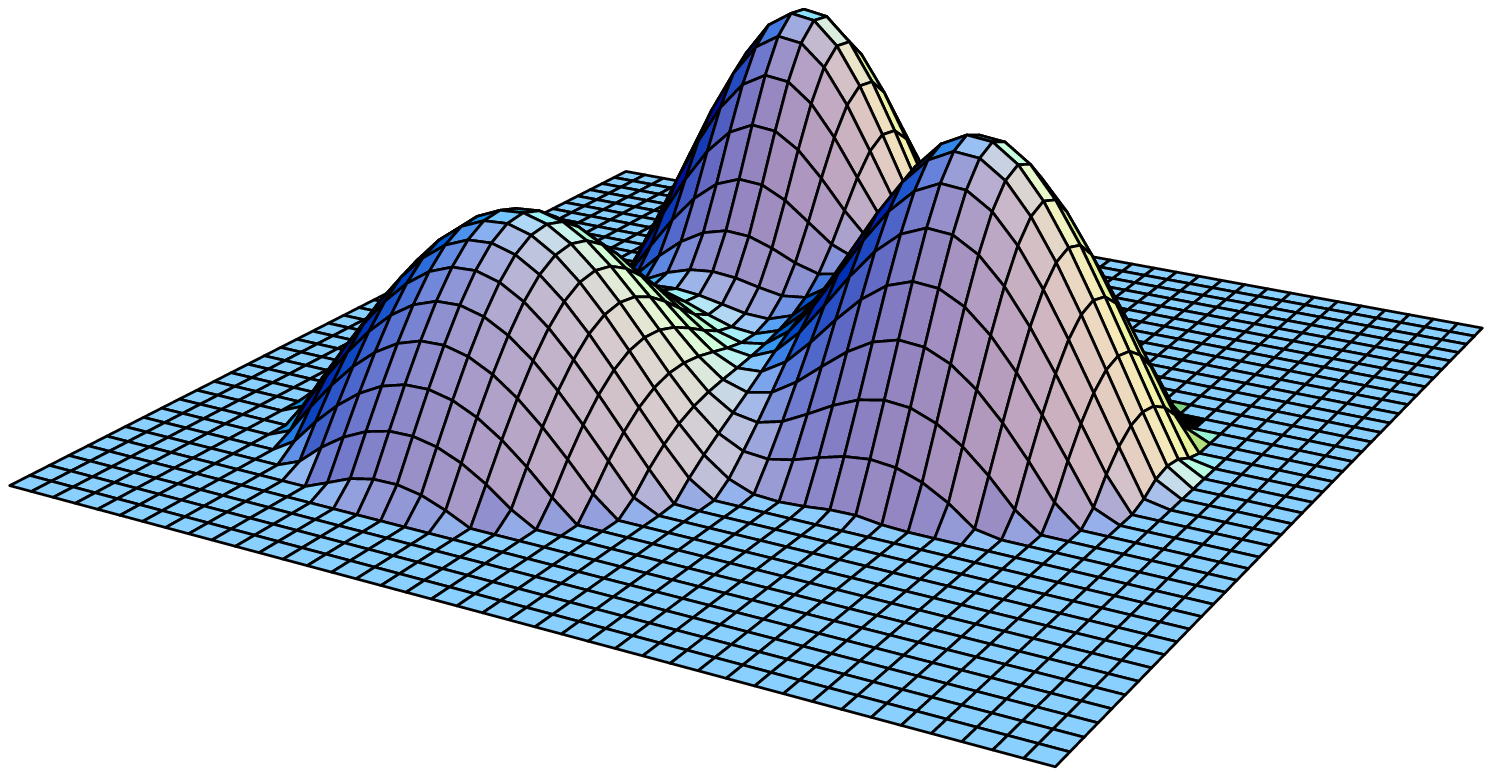}
\includegraphics{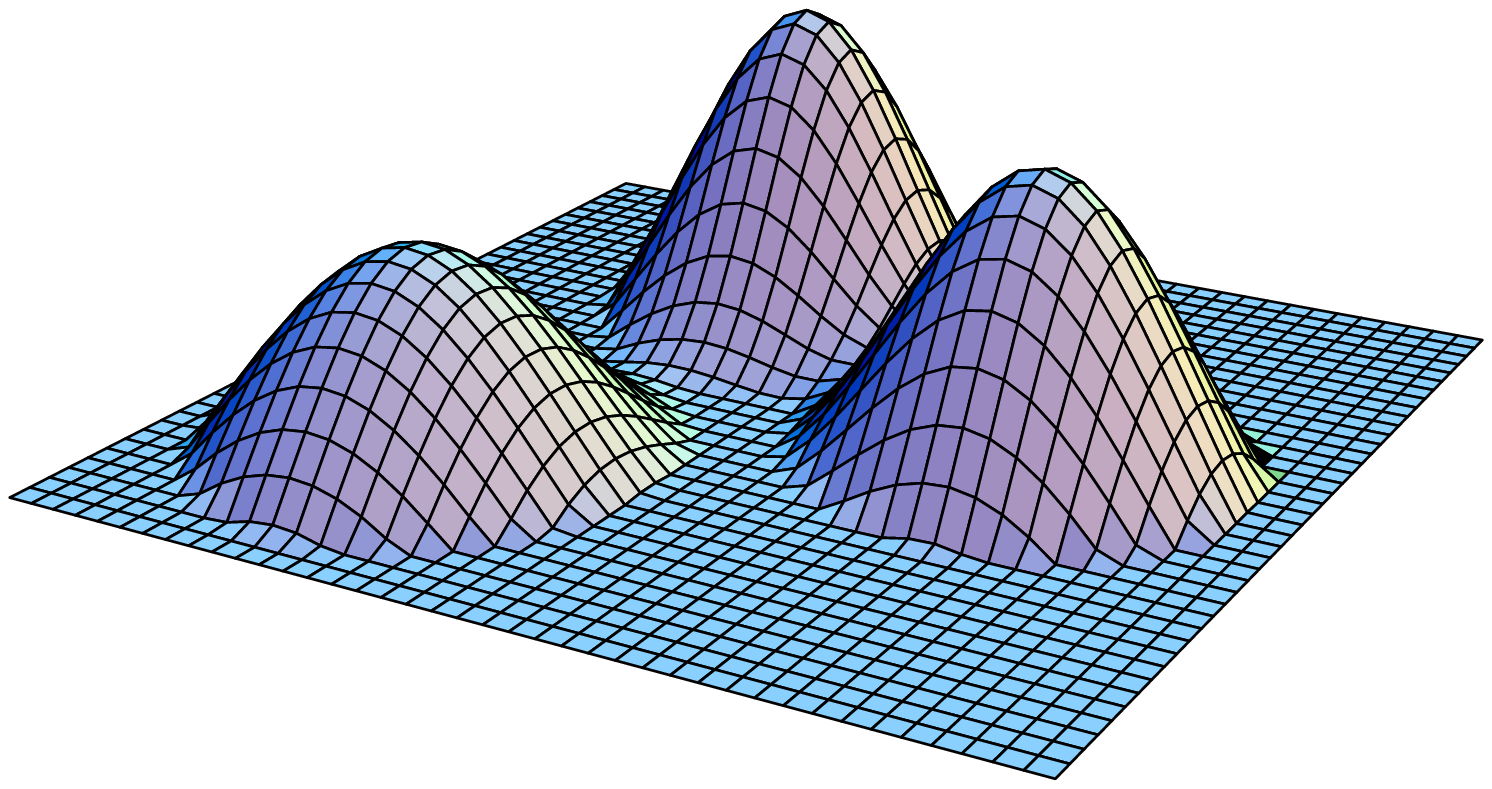}
\caption{Action densities for the $SU(3)$ caloron at $x_0=0$ in the plane 
defined by the centers of the three constituents for $1/\cT=1.5,3$ and 4 
(increasing temperature from top to bottom). We choose mass parameters 
$(\nu_1,\nu_2,\nu_3)=(0.4,0.35,0.25)$, implemented by $(\mu_1,\mu_2,\mu_3)=
(-17/60,-2/60,19/60)$. The constituents are located at $\vec y_1=(-\half,\half,
0)$, $\vec y_2=(0,\half,0)$ and $\vec y_3=(\half, -\quarter,0)$, in units of 
$\cT$. The profiles are given on equal logarithmic scales, cut-off at an 
action density below $1/e$.}
\end{figure}

When the lumps are far apart, they do not deform each other and become 
spherically symmetric. Since the solution is self-dual, the constituents 
have to be basic BPS monopoles. This can be proven by carefully analysing 
\refeq{adpsi} for the limit where $r_m\ll r_l$ for all $l\neq m$, in which 
case the action density approaches $-\half\partial_\mu^2\partial_\nu^2\log
[\sinh(2\pi\nu_m r_m)/r_m]$. This is precisely the behaviour of the BPS 
monopole~\cite{Ros}. The other constituents need not be well-separated from 
each other for the above argument to hold. In particular sending the 
$m^{\rm th}$ constituent to infinity (i.e. $|\vec y_m|\rightarrow\infty$) 
suffices to make the caloron static. What remains are $n-1$ monopole 
constituents with a combined magnetic charge opposite to the magnetic 
charged of the $m^{\rm th}$ constituent monopole that has been removed. 
As the solution is static in this limit one is left with an $SU(n)$ BPS 
monopole. Indeed, for $|\vec y_m|\rightarrow\infty$ we see from the 
solution of the Nahm equation, \refeq{abelnahm}, that $\hat A_i(z)$ lives 
on an interval, rather than on the circle, as is appropriate for the $SU(n)$ 
monopole~\cite{NahAll}. One readily obtains the energy density of this 
monopole by taking the limit $|\vec y_m|\rightarrow\infty$ in \refeq{psi}, 
verifying that it decays as $1/|\vec x|^4$, as opposed to $1/|\vec x|^6$ 
without removing the $m^{\rm th}$ constituent.

Our results have been derived for the case of maximal symmetry breaking, 
$\mu_m\neq\mu_{m+1}$. The situation of non-maximal symmetry breaking 
corresponds to a constituent obtaining zero mass, $\nu_m=0$. In that 
case its center of mass radius {\em drops out} of \refeq{psi}, using
\beq 
\left(\!\!\!\bea{cc}r_m&|\vec y_m\!-\!\vec y_{m+1}|\\0&r_{m+1}\eea\!\!\!\right)
\frac{1}{r_m}\left(\!\!\!\bea{cc}r_{m-1}&|\vec y_m\!-\!\vec y_{m-1}|\\0&r_m
\eea\!\!\!\right)=\left(\!\!\!\bea{cc}r_{m-1}&|\vec y_m\!-\!\vec y_{m+1}|+
|\vec y_m\!-\!\vec y_{m-1}|\\0&r_{m+1}\eea\!\!\!\right).
\eeq
This was also observed for $SU(2)$, in which case non-maximal symmetry 
breaking corresponds to a trivial Polyakov loop, $\pl=\pm1$, and the
solution becomes that of Harrington and Shepard~\cite{HarShe}. Hence
our formula for the action density should also be valid for non-maximal
symmetry breaking.

Although our formalism can be extended easily to higher topological 
charges~\cite{LeeYi2}, the appropriate Nahm equation (i.e. solving the 
quadratic ADHM constraint) becomes a non-abelian problem, and finding 
solutions requires more powerful tools. Nevertheless, it is interesting 
to note that it is natural to conjecture that $k$ instantons (\ie\ an 
instanton of charge $k$) can be built from $kn$ monopoles, since each 
instanton can be considered as being built from $n$ BPS monopoles. The 
monopole constituents are only well separated when $\cT$ is small, where 
the $4kn$ instanton parameters can be interpreted as $3kn$ positions and 
$kn$ phases (including $\exp(2\pi i\xi_0/\cT)$). We will not speculate 
further on these matters here, but want to emphasise that the monopole 
constituent picture has some interesting phenomenological implications 
for the description of the long distance properties of QCD, discussed 
in detail in ref.~\cite{NPB}, and which will be the subject of further 
investigations.

\section*{Acknowledgements}
TCK was supported by a grant from the FOM/SWON Association for Mathematical
Physics.

\end{document}